\documentclass[aps,twocolumn,groupedaddress,showpacs]{revtex4}%
\usepackage{amsmath}
\usepackage{amsfonts}
\usepackage{amssymb}
\usepackage{graphicx}%
\begin{document}
\title{Classical $O(N)$ nonlinear sigma model on the half line: \\a study on consistent Hamiltonian description}
\author{Wenli He}
\affiliation{Institute of Modern Physics, Northwest University, Xian 710069, China}
\author{Liu Zhao\footnote{Correspondence author, e-mail: lzhao@nwu.edu.cn}}
\affiliation{Institute of Modern Physics, Northwest University, Xian 710069, China}

\begin{abstract}
The problem of consistent Hamiltonian structure for $O(N)$ nonlinear sigma
model in the presence of five different types of boundary conditions is
considered in detail. For the case of Neumann, Dirichlet and the mixture of
these two types of boundaries, the consistent Poisson brackets are constructed
explicitly, which may be used, e.g. for the construction of current algebras
in the presence of boundary. While for the mixed boundary conditions and the
mixture of mixed and Dirichlet boundary conditions, we prove that there is no
consistent Poisson brackets, showing that the mixed boundary conditions are
incompatible with all nontrivial subgroups of $O((N)$.

\end{abstract}
\maketitle

\vspace{1cm}

\section{Introduction}

Field theories with boundaries have been attracting the attention of
theoretical physicists for a number of reasons, especially from the quantum
point of view. For example, the existence of boundaries is responsible for the
Casimir effect and surface phenomena, fundamental excitations in the bulk may
have interesting behavior when scattered off the boundaries \cite{Ghoshal:NLS}%
, and sometimes boundary bound state might appear, etc. Another important
aspect of boundaries appear in the study of string theory, where they are used
to distinguish different types of string theories and are also regarded as the
reason for the occurrence of noncommutativity on the D-branes.

The introduction of boundary interactions into the Lagrangian also causes some
problem at the classical level, since the boundary conditions would in general
spoil the naive Poisson structure. In order to describe classical field
theories with boundaries as consistent Hamiltonian systems, many authors
prefer to use the Dirac method for treating constraints
\cite{Dirac1,Dirac2,Dirac3}. However, as pointed out in \cite{Zhao}, the
direct application of Dirac method in boundary systems has some problems,
mostly due to the fact that boundary conditions regarded as constraints have
functional measure $0$ in the space of fields. To overcome these problems,
some authors prefer to use modified versions of Dirac method
\cite{Deriglazov1}, or first turn the field theories with boundaries into
mechanical systems with infinite many degrees of freedom by use of either
Fourier mode expansion or lattice approximation and then use the Dirac
approach \cite{Sheikh-Jabbari1,Chu1,Chu2,Arfaei}. Other methods, including
symplectic quantization \cite{Braga,Faddeev1} and lightcone quantization
\cite{Maeno}, are also used to treat boundary problems. However, none of the
above mentioned methods is applicable systematically to all field theories
with boundaries. The approach which works for one particular model with ease
may become very cumbersome, or even completely inapplicable to use for
another. In \cite{Zhao}, we proposed a novel method for treating the boundary
constraints. Our method is based on a very simple idea, i.e. the principle of
locality: since the boundary conditions are constraints only at the
boundaries, they should modify the naive Poisson structure only at the
boundaries. By directly modifying the naive Poisson brackets at the boundaries
with some test operators and checking the compatibility with boundary
constraints, we can obtain conditions to determine the test operators. This
method is used in the subsequent works \cite{He} and \cite{Zhao2} to study the
problem of open string quantization in background NS-NS $B$-field, and is
proven to be very powerful and easy to use.

In this Letter, we are aimed at using the method of \cite{Zhao} to study the
problem of consistent Hamiltonian description for $O(N)$ nonlinear sigma model
in the presence of integrable boundary conditions \cite{Moriconi3,Moriconi2}.
Besides getting more concrete examples for the application of our method,
there are more direct motivations to study the Hamiltonian description for
this model. In the literatures, $O(N)$ nonlinear sigma model is often taken as
a typical model of field theories with certain nice geometrical properties
\cite{Braaten}, it is also a frequently used toy model for stimulating
nonabelian gauge theories \cite{Eichenherr,Cremmer,Golo,D'Adda}, and a
theoretical laboratory for exploring Poisson-Lie geometry and current algebras
\cite{Forger}. In a number of problems in statistical physics, condensed
matter systems \cite{LeClair,Wen} and/or high energy physics, e.g. quantum
antiferromagnetism, large $N$ behavior and asymptotic freedom in strong
interactions, $O(N)$ nonlinear sigma model is often found to be a simplified
version of the underlying field theoretic description. Another area in which
nonlinear sigma model found important applications is string theory. There the
model is often used to describe D-brane dynamics in curved backgrounds
\cite{Barabanschikov}. The exact integrability of $O(N)$ nonlinear sigma model
on the half line \cite{Moriconi1,Moriconi2,Moriconi3,Mourad} provides more
direct motivations for the mathematical physicists to study this theory. In
this respect, the study of consistent Hamiltonian description of $O(N)$
nonlinear sigma model is quite essential, because the quantum analysis on the
factorized scattering in the bulk as well as off the boundary based on quantum
inverse scattering method needs semiclassical support, for which the classical
Hamiltonian description of the model is a starting point. Even from a pure
classical integrable system point of view, a consistent Hamiltonian
description is still a key structure because it is needed to prove that the
integrals of motion are pairwise in involution under the correct Poisson
brackets. However, to our knowledge, a systematical analysis on the
Hamiltonian structure of the $O(N)$ nonlinear sigma model in the presence of
integrable boundary conditions is still not undertaken, at least in the form
we shall present. That's why we start our analysis from now on.

\section{The model on the half line}

The action for $O(N)$ nonlinear sigma model in $(1+1)$-spacetime dimensions reads%

\begin{equation}
S=\frac{1}{2}\int d^{2}x\left[  \partial_{\mu}\mathbf{n}^{T}\cdot\partial
^{\mu}\mathbf{n}+\omega(\mathbf{n}^{T}\cdot\mathbf{n}-1)\right]  ,
\label{action}%
\end{equation}
where the field $\mathbf{n}=(n_{1},n_{2},n_{3},\cdots,n_{N})^{T}$ obey the
$O(N)$ condition $\mathbf{n}^{T}\cdot\mathbf{n}=1$, thanks to the Lagrangian
multiplier $\omega$. We use the superscript $T$ to represent matrix transpose.
The spacetime metric we adopt is $(\eta_{\mu\nu})=diag(1,-1)$, and summation
over repeated indices is assumed throughout.

The variation of (\ref{action}) with respect to $\mathbf{n}$ leads to the
equation of motion
\begin{equation}
\partial_{\mu}\partial^{\mu}\mathbf{n}^{T}-\omega\mathbf{n}^{T}=0. \label{3}%
\end{equation}
By use of the $O(N)$ condition $\mathbf{n}^{T}\cdot\mathbf{n}=1$£¬(\ref{3})
can be rewritten as
\begin{equation}
\partial_{\mu}\partial^{\mu}\mathbf{n}+\left(  \partial_{\mu}\mathbf{n}%
^{T}\cdot\partial^{\mu}\mathbf{n}\right)  \mathbf{n}=0. \label{5}%
\end{equation}

In the Hamiltonian description, the fundamental dependent variables are the
fields (\textquotedblleft canonical coordinates\textquotedblright) and their
conjugate momenta. The conjugate momenta in the bulk are defined as
\begin{equation}
\pi_{i}\equiv\frac{\delta\mathcal{L}_{B}}{\delta(\partial_{t}n^{i})}%
=\partial_{t}n_{i}. \label{6}%
\end{equation}
Since the $O(N)$ condition $\mathbf{n}^{T}\cdot\mathbf{n}=1$ is a constraint,
the correct Poisson brackets for the fields $n_{i}$ and the conjugate momenta
$\pi_{i}$ must be obtained by use of the standard Dirac method. The results
read
\begin{equation}
\{n_{i}(x),n_{j}(y)\}=0, \label{7}%
\end{equation}%
\begin{equation}
\{n_{i}(x),\pi_{j}(y)\}=(\delta_{ij}-n_{i}n_{j})\delta(x-y), \label{8}%
\end{equation}%
\begin{equation}
\{\pi_{i}(x),\pi_{j}(y)\}=(\pi_{i}n_{j}-n_{i}\pi_{j})\delta(x-y). \label{9}%
\end{equation}
This finishes the description of the model in the bulk.

In the presence of a boundary, the form of the Lagrangian is kept unchanged,
but the spacial integration in (\ref{action}) is restricted on the half line
$x\in\lbrack0,\infty)$. Several types of boundary conditions are
\emph{claimed} to be integrable in the literatures \cite{Moriconi3,Moriconi2}.
They are (i) Neumann boundary conditions along all target space directions,
i.e. $\partial_{x}n_{i}|_{x=0}=0$, $i=1,\cdots,N$. We denote this set of
boundary conditions as (AN) (i.e. all Neumann); (ii) Dirichlet boundary
conditions along all target space directions, i.e. $\partial_{t}n_{i}%
|_{x=0}=0$, $i=1,\cdots,N$. This set of boundary conditions is denoted as (AD)
(i.e. all Dirichlet); (iii) a mixture of Neumann and Dirichlet boundary
conditions, i.e. $\partial_{x}n_{i}|_{x=0}=0$ for $i=1,\cdots,p$ and
$\partial_{t}n_{i}|_{x=0}=0$ for $i=p+1,\cdots,N$. This set of boundary
conditions is denoted as (ND) (i.e. mixed Neumann and Dirichlet); (iv) mixed
boundary conditions along all target space directions, i.e. $(\partial
_{x}n_{i}+M_{ij}\partial_{t}n_{j})|_{x=0}=0$ for $i=1,\cdots,N$, where $M$ is
a real invertible antisymmetric matrix of the form%
\begin{equation}
M=g_{1}(i\sigma^{2})\oplus g_{2}(i\sigma^{2})\oplus\cdots\oplus g_{K}%
(i\sigma^{2}), \label{M}%
\end{equation}
in which $\sigma^{2}$ is the second Pauli matrix, $g_{1}$ through $g_{K}$ are
free parameters (boundary coupling constants). Notice that this type of
boundary conditions is only possible for even $N=2K$, because otherwise $M$
cannot not be invertible. This set of boundary conditions is actually not
found in \cite{Moriconi3,Moriconi2}, but is a simple generalization of the
non-diagonal boundary conditions proposed there (the non-diagonal boundary
condition in \cite{Moriconi3,Moriconi2} contains only one $i\sigma^{2}$
block). We shall refer to this set of boundary conditions as (AM) (all mixed);
(v) a mixture of mixed and Dirichlet boundary conditions, i.e. $(\partial
_{x}n_{i}+\mathcal{M}_{ij}\partial_{t}n_{j})|_{x=0}=0$ for $i,j=1,\cdots,p$
$(p=2K)$ and $\partial_{t}n_{i}=0$ for $i=p+1,\cdots,N$, where $\mathcal{M}$
is given as the $M$ in (\ref{M}). This last set of boundary conditions is
denoted as (MD). It has been mentioned in \cite{Moriconi3,Moriconi2} that the
mixture of mixed and Neumann boundary conditions (MN) is not integrable, at
least on the quantum level. We thus exclude this case from our consideration.

To put things together, it is useful to introduce another matrix%
\begin{equation}
W=\left(
\begin{array}
[c]{cc}%
\mathcal{W} & \\
& 0_{N-p}%
\end{array}
\right)  , \label{W}%
\end{equation}
in which $\mathcal{W=M}^{-1}$, the inverse of $\mathcal{M}$. Then the MD
boundary conditions can be written in the following unified form,%
\begin{equation}
(\partial_{t}n_{i}+W_{ij}\partial_{x}n_{j})|_{x=0}=0,\quad i=1,\cdots,N.
\label{4pp}%
\end{equation}
Moreover, the form of (\ref{4pp}) also contains the other 4 types of boundary
conditions mentioned above as special degenerated cases, if we allow the
matrix $\mathcal{W}$ to take different forms. Concretely, (\ref{4pp}) will be
reduced into AD boundaries for $p=0$, into AM boundaries for $p=N=2K$ and
$W=M^{-1}$; for generic $p$ with $\mathcal{W}$ diagonal and all $\mathcal{W}%
_{ii}\rightarrow\infty$, (\ref{4pp}) will be reduced into ND boundaries; and
for $p=N$ with $\mathcal{W}$ diagonal and all $\mathcal{W}_{ii}\rightarrow
\infty$, it will be reduced into AN boundaries. We therefore will take
(\ref{4pp}) as the starting point for our analysis.

It should be remarked that, in the presence of the boundary conditions
(\ref{4pp}), there is some ambiguity in the definition of canonical conjugate
momenta, because the mixed boundary conditions can be realized via variational
principle by adding a boundary term to the action which contains $\partial
_{t}n_{i}$. The additional boundary term makes the canonical momenta defined
as variations of the \emph{complete Lagrangian} $\mathcal{L}$ with respect to
the time derivatives of the fields $n_{j}$ differ from those defined as
variations of the \emph{bulk Lagrangian} $\mathcal{L}_{B}$. For our purpose,
it is more convenient to stick to the bulk momenta $\pi_{i}$, because there is
already a set of known Poisson brackets (\ref{7})-(\ref{9}) which can be taken
as the basis of our analysis. Using the phase space variables $n_{i}$ and
$\pi_{i}$, we can rewrite the boundary conditions (\ref{4pp}) as%
\begin{equation}
(\pi_{i}+W_{ij}\partial_{x}n_{j})|_{x=0}=0. \label{4p}%
\end{equation}
It can be seen that, since the boundary conditions (\ref{4p}) identify
$\partial_{x}n_{i}$ with some specific linear combination of $\pi_{i}$, the
Poisson brackets (\ref{7})-(\ref{9}) would no longer hold. In the next
section, we shall try to construct consistent Poisson brackets which are
compatible with (\ref{4p}). However, it will turn out that only for AD, AN and
ND boundaries we can make a success. For AM and MD boundaries we can find no
consistent Poisson brackets, which indicates that the mixed boundary
conditions are not allowed for $O(N)$ nonlinear sigma model.

\section{Boundary constraints and general compatibility conditions}

Following the method of \cite{Zhao}, the very first step in getting consistent
modifications of the Poisson brackets (\ref{7})-(\ref{9}) would be introducing
the \emph{boundary constraints}%

\begin{equation}
G_{i}\equiv\int_{0}^{\infty}dx\delta(x)(\pi_{i}+W_{ij}\partial_{x}n_{j}%
)\simeq0. \label{10}%
\end{equation}
This is just another way of writing the boundary conditions (\ref{4p}), in
which the $\delta$-function is a slightly regularized one \cite{Zhao},
satisfying $\int_{0}^{\infty}dx\delta(x)=1$.

Since the constraints $G_{i}$ are strong zeros beyond the boundary at $x=0$,
it is tempting to think that there is no need to modify (\ref{7})-(\ref{9})
except at $x=0$, and it was indeed so in the cases of \cite{Zhao,He,Zhao2}.
However, at this point, we would prefer to keep things as general as possible.
Therefore, assuming that the consistent bulk Poisson brackets take the form%
\begin{align}
\{n_{i}(x),n_{j}(y)\}  &  =\mathfrak{A}_{ij}(n,\pi)\delta(x-y),\nonumber\\
\{n_{i}(x),\pi_{j}(y)\}  &  =\mathfrak{B}_{ij}(n,\pi)\delta(x-y),\nonumber\\
\{\pi_{i}(x),\pi_{j}(y)\}  &  =\mathfrak{C}_{ij}(n,\pi)\delta(x-y)
\label{PBulk}%
\end{align}
and adding boundary modifications, the most general form for the potential
consistent Poisson brackets will be
\begin{equation}
\{n_{i}(x),n_{j}(y)\}_{M}=\mathfrak{A}_{ij}(n,\pi)\delta(x-y)+\mathcal{A}%
_{ij}\delta(x+y), \label{11}%
\end{equation}%
\begin{equation}
\{n_{i}(x),\pi_{j}(y)\}_{M}=\mathfrak{B}_{ij}(n,\pi)\delta(x-y)+\mathcal{B}%
_{ij}\delta(x+y), \label{12}%
\end{equation}%
\begin{equation}
\{\pi_{i}(x),\pi_{j}(y)\}_{M}=\mathfrak{C}_{ij}(n,\pi)\delta(x-y)+\mathcal{C}%
_{ij}\delta(x+y), \label{13}%
\end{equation}
where the suffix $M$ denotes modified Poisson brackets, $\mathfrak{A,B,C}$ are
some\emph{ known} functions in the phase space with $\mathfrak{A}_{ij}$ and
$\mathfrak{C}_{ij}$ antisymmetric in $i\leftrightarrow j$, and $\mathcal{A}$,
$\mathcal{B}$, $\mathcal{C}$ are some operators acting on the variable $y$
which are yet to be determined by consistency requirements. Since the Poisson
brackets are antisymmetric, the operators $\mathcal{A}_{ij}$ and
$\mathcal{C}_{ij}$ \emph{must }also\emph{ be antisymmetric} in
$i\leftrightarrow j$.

At first sight, it may look strange that we assume the odd form (\ref{PBulk})
for the bulk Poisson brackets rather than use (\ref{7})-(\ref{9}) directly.
The reason for this will be clear in the next section when we try to find
solutions for the compatibility conditions which we now derive.

In order to determine the values of $\mathcal{A}$, $\mathcal{B}$ and
$\mathcal{C}$, we first apply the compatibility conditions%
\begin{align}
\{G_{i},n_{j}(y)\}_{M}  &  =0,\label{comp:1}\\
\{G_{i},\pi_{j}(y)\}_{M}  &  =0. \label{comp:2}%
\end{align}
Straightforward calculations yield
\begin{align}
&  \{G_{i},n_{j}(y)\}_{M}\nonumber\\
&  \hspace{2mm}=\int_{0}^{\infty}dx\delta(x)\{\pi_{i}+W_{ik}\partial_{x}%
n_{k},n_{j}(y)\}_{M}\nonumber\\
&  \hspace{2mm}=\int_{0}^{\infty}dx\delta(x)\left(  \{\pi_{i},n_{j}%
(y)\}_{M}+\{W_{ik}\partial_{x}n_{k},n_{j}(y)\}_{M}\right) \nonumber\\
&  \hspace{2mm}=\int_{0}^{\infty}dx\delta(x)\left[  -\mathfrak{B}_{ji}%
(n,\pi)\delta(x-y)-\mathcal{B}_{ji}\delta(x+y)\right. \nonumber\\
&  \hspace{2mm}\hspace{2mm}+\left.  W_{ik}\partial_{x}\{\mathfrak{A}%
_{kj}(n,\pi)\delta(x-y)+\mathcal{A}_{kj}\delta(x+y)\}\right] \nonumber\\
&  \hspace{2mm}=-\left[  \left(  \mathfrak{A-}\mathcal{A}\right)
W\partial_{y}+(\mathfrak{B}+\mathcal{B})\right]  _{ji}\delta(y), \label{14}%
\end{align}%
\begin{align}
&  \{G_{i},\pi_{j}(y)\}_{M}\nonumber\\
&  \,=\int_{0}^{\infty}dx\delta(x)\{\pi_{i}+W_{ik}\partial_{x}n_{k},\pi
_{j}(y)\}_{M}\nonumber\\
&  =\int_{0}^{\infty}dx\delta(x)\left(  \{\pi_{i},\pi_{j}(y)\}_{M}%
+\{W_{ik}\partial_{x}n_{k},\pi_{j}(y)\}_{M}\right) \nonumber\\
&  =\int_{0}^{\infty}dx\delta(x)\left[  \mathfrak{C}_{ij}(n,\pi)\delta
(x-y)+\mathcal{C}_{ij}\delta(x+y)\right. \nonumber\\
&  +\left.  W_{ik}\partial_{x}\left(  \mathfrak{B}_{ij}(n,\pi)\delta
(x-y)+\mathcal{B}_{ij}\delta(x+y)\right)  \right] \nonumber\\
&  \,=[\mathfrak{C}+\mathcal{C}-W(\mathfrak{B}-\mathcal{B})\partial_{y}%
]_{ij}\delta(y), \label{15p}%
\end{align}
where $\mathbf{\pi}=(\pi_{1},\pi_{2},\pi_{3},\cdots,\pi_{N})^{T}$. Comparing
(\ref{14}), (\ref{15p}) to the compatibility conditions (\ref{comp:1}) and
(\ref{comp:2}), we get the following equation for the operators $\mathcal{A}$,
$\mathcal{B}$ and $\mathcal{C}$,
\begin{equation}
\left(  \mathfrak{A-}\mathcal{A}\right)  W\partial_{y}+(\mathfrak{B}%
+\mathcal{B})=0, \label{16}%
\end{equation}%
\begin{equation}
\mathfrak{C}+\mathcal{C}-W(\mathfrak{B}-\mathcal{B})\partial_{y}=0. \label{17}%
\end{equation}

The compatibility between the test Poisson brackets and the boundary
constraints do not provide the complete set of compatibility conditions for
the operators $\mathcal{A}$, $\mathcal{B}$ and $\mathcal{C}$. In order that
the test Poisson brackets (\ref{11})-(\ref{13}) be fully consistent, they are
also required to satisfy Jacobi identities. For the canonical variables
$n_{i},\pi_{j}$, there are totally $4$ different types of Jacobi identities to
check, i.e. the ones for $\{n_{i},n_{j},n_{k}\}$, $\{n_{i},n_{j},\pi_{k}\}$,
$\{n_{i},\pi_{j},\pi_{k}\}$ and $\{\pi_{i},\pi_{j},\pi_{k}\}$ respectively.
These identities hold identically beyond the boundary, because the bulk
Poisson brackets (\ref{PBulk}) are already consistent before implementing the
boundary constraints. Therefore, what we need to check are only the Jacobi
identities at the boundary. Using (\ref{11})-(\ref{13}), we get from the above
mentioned Jacobi identities the following equations,%
\begin{align}
&  \frac{\delta\left(  \mathfrak{A+}\mathcal{A}\right)  _{ij}}{\delta n_{m}%
}\left(  \mathfrak{A+}\mathcal{A}\right)  _{mk}-\frac{\delta\left(
\mathfrak{A+}\mathcal{A}\right)  _{ij}}{\delta\pi_{m}}\left(  \mathfrak{B}%
+\mathcal{B}\right)  _{km}\nonumber\\
&  \,+\frac{\delta\left(  \mathfrak{A+}\mathcal{A}\right)  _{jk}}{\delta
n_{m}}\left(  \mathfrak{A+}\mathcal{A}\right)  _{mi}-\frac{\delta\left(
\mathfrak{A+}\mathcal{A}\right)  _{jk}}{\delta\pi_{m}}\left(  \mathfrak{B}%
+\mathcal{B}\right)  _{im}\nonumber\\
&  \,+\frac{\delta\left(  \mathfrak{A+}\mathcal{A}\right)  _{ki}}{\delta
n_{m}}\left(  \mathfrak{A+}\mathcal{A}\right)  _{mj}-\frac{\delta\left(
\mathfrak{A+}\mathcal{A}\right)  _{ki}}{\delta\pi_{m}}\left(  \mathfrak{B}%
+\mathcal{B}\right)  _{jm}=0,\label{Jac:1}%
\end{align}%
\begin{align}
&  \frac{\delta\left(  \mathfrak{A+}\mathcal{A}\right)  _{ij}}{\delta n_{m}%
}\left(  \mathfrak{B}+\mathcal{B}\right)  _{mk}+\frac{\delta\left(
\mathfrak{A+}\mathcal{A}\right)  _{ij}}{\delta\pi_{m}}\left(  \mathfrak{C}%
+\mathcal{C}\right)  _{mk}\nonumber\\
&  +\frac{\delta\left(  \mathfrak{B}+\mathcal{B}\right)  _{jk}}{\delta n_{m}%
}\left(  \mathfrak{A+}\mathcal{A}\right)  _{mi}-\frac{\delta\left(
\mathfrak{B}+\mathcal{B}\right)  _{jk}}{\delta\pi_{m}}\left(  \mathfrak{B}%
+\mathcal{B}\right)  _{im}\nonumber\\
&  -\frac{\delta\left(  \mathfrak{B}+\mathcal{B}\right)  _{ik}}{\delta n_{m}%
}\left(  \mathfrak{A+}\mathcal{A}\right)  _{mj}+\frac{\delta\left(
\mathfrak{B}+\mathcal{B}\right)  _{ik}}{\delta\pi_{m}}\left(  \mathfrak{B}%
+\mathcal{B}\right)  _{jm}=0,\label{Jac:2}%
\end{align}%
\begin{align}
&  \frac{\delta\left(  \mathfrak{B}+\mathcal{B}\right)  _{ij}}{\delta n_{m}%
}\left(  \mathfrak{B}+\mathcal{B}\right)  _{mk}+\frac{\delta\left(
\mathfrak{B}+\mathcal{B}\right)  _{ij}}{\delta\pi_{m}}\left(  \mathfrak{C}%
+\mathcal{C}\right)  _{mk}\nonumber\\
&  +\frac{\delta\left(  \mathfrak{C}+\mathcal{C}\right)  _{jk}}{\delta n_{m}%
}\left(  \mathfrak{A+}\mathcal{A}\right)  _{mi}-\frac{\delta\left(
\mathfrak{C}+\mathcal{C}\right)  _{jk}}{\delta\pi_{m}}\left(  \mathfrak{B}%
+\mathcal{B}\right)  _{im}\nonumber\\
&  -\frac{\delta\left(  \mathfrak{B}+\mathcal{B}\right)  _{ik}}{\delta n_{m}%
}\left(  \mathfrak{B}+\mathcal{B}\right)  _{mj}-\frac{\delta\left(
\mathfrak{B}+\mathcal{B}\right)  _{ik}}{\delta\pi_{m}}\left(  \mathfrak{C}%
+\mathcal{C}\right)  _{mj}=0,\label{Jac:3}%
\end{align}
and%
\begin{align}
&  \frac{\delta\left(  \mathfrak{C}+\mathcal{C}\right)  _{ij}}{\delta n_{m}%
}\left(  \mathfrak{B}+\mathcal{B}\right)  _{mk}\,+\frac{\delta\left(
\mathfrak{C}+\mathcal{C}\right)  _{ij}}{\delta\pi_{m}}\left(  \mathfrak{C}%
+\mathcal{C}\right)  _{mk}\nonumber\\
&  \,+\frac{\delta\left(  \mathfrak{C}+\mathcal{C}\right)  _{jk}}{\delta
n_{m}}\left(  \mathfrak{B}+\mathcal{B}\right)  _{mi}\,+\frac{\delta\left(
\mathfrak{C}+\mathcal{C}\right)  _{jk}}{\delta\pi_{m}}\left(  \mathfrak{C}%
+\mathcal{C}\right)  _{mi}\nonumber\\
&  \,+\frac{\delta\left(  \mathfrak{C}+\mathcal{C}\right)  _{ki}}{\delta
n_{m}}\left(  \mathfrak{B}+\mathcal{B}\right)  _{mj}\,+\frac{\delta\left(
\mathfrak{C}+\mathcal{C}\right)  _{ki}}{\delta\pi_{m}}\left(  \mathfrak{C}%
+\mathcal{C}\right)  _{mj}=0.\label{Jac:4}%
\end{align}
Once the equations (\ref{Jac:1})-(\ref{Jac:4}) are satisfied, the Jacobi
identities for any functions on the phase space will hold consistently,
because $n_{i},\pi_{j}$ form a basis for the phase space of the model.
Therefore we conclude that the system of equations (\ref{16})-(\ref{Jac:4}) is
the complete set of conditions which the operators $\mathcal{A}$,
$\mathcal{B}$, $\mathcal{C}$ must obey. As long as a solution \{$\mathcal{A}$,
$\mathcal{B}$, $\mathcal{C}$\} to the above system of operator equations is
found, we will get a consistent Hamiltonian description for $O(N)$ nonlinear
sigma model with the boundary conditions (\ref{4pp}). However, since the
system of equations (\ref{16})-(\ref{Jac:4}) is over determined, the existence
of a solution is not guaranteed in general. When no solution to (\ref{16}%
)-(\ref{Jac:4}) can be found, the nonexistence of a solution should be
considered as a signature that the corresponding boundary conditions are
incompatible with the bulk dynamics. In the next section, we shall show that
the AM and MD boundaries belong to this forbidden class of boundaries. The
other three types of boundaries, i.e. AD, AN and ND boundaries, will all give
rise to consistent solutions to the compatibility equations (\ref{16}%
)-(\ref{Jac:4}).

\section{Consistent Poisson brackets}

In this section, we shall try to find explicit solutions for the system of
equations (\ref{16})-(\ref{Jac:4}) under each of the five different types of
boundary conditions mentioned earlier. The basic strategy in getting these
special solutions is like this: we shall first try to get solutions to the
relatively simpler equations (\ref{16}), (\ref{17}) and then check that they
are consistent with the rest equations, (\ref{Jac:1})-(\ref{Jac:4}). All
solutions to the system of equations (\ref{16})-(\ref{Jac:4}) can in principle
be obtained in this manner.

\subsection{$O(N)$ symmetric boundaries AD and AN}

The first types of boundaries we shall consider are the AD and AN boundaries,
which can be easily seen to preserve the complete $O(N)$ symmetry of the
model. We shall treat both of these two types of boundary conditions in a
unified way by use of the boundary constraints (\ref{10}) and requiring $p$ to
be either $0$ or equal to $N$. Doing so we are seemingly to be considering the
AD, AN and AM boundaries in a unified manner. However, it will be clear
shortly that the AM case is distinguished from the AD and AN cases, because AM
is actually symmetry breaking.

Now let us look at the equations (\ref{16}), (\ref{17}) in more detail. Since
we are now considering symmetry preserving boundaries, there is no problem to
identify the bulk Poisson brackets (\ref{PBulk}) with (\ref{7})-(\ref{9}),
i.e. to choose $\mathfrak{A}_{ij}=0,\mathfrak{B}_{ij}=\delta_{ij}-n_{i}n_{j}$
and $\mathfrak{C}_{ij}=\pi_{i}n_{j}-\pi_{j}n_{i}$. Then (\ref{16}), (\ref{17})
will become%
\begin{align}
&  \mathcal{A}_{im}W_{mj}\partial_{y}-\left(  I-\mathbf{n}\cdot\mathbf{n}%
^{T}+\mathcal{B}\right)  _{ij}=0,\label{16p}\\
&  \left(  \mathbf{\pi}\cdot\mathbf{n}^{T}-\mathbf{n}\cdot\mathbf{\pi}%
^{T}+\mathcal{C}\right)  _{ij}-W_{im}\left(  I-\mathbf{n}\cdot\mathbf{n}%
^{T}-\mathcal{B}\right)  _{mj}\partial_{y}=0. \label{17p}%
\end{align}
To solve the last two equations, we need to consider three different cases,
i.e. a) $p=0$ or effectively $W=0$; b) $p=N$ with $W$ diagonal and
$W_{ii}\rightarrow\infty$ for all $i$; c) $p=N=2K$ and $W=M^{-1}$ with $M$
given in (\ref{M}). In case a) we get from (\ref{16p}) and (\ref{17p}) the
result
\begin{align*}
\mathcal{B}_{ij}  &  =-\left(  I-\mathbf{n}\cdot\mathbf{n}^{T}\right)
_{ij},\\
\mathcal{C}_{ij}  &  =-\left(  \mathbf{\pi}\cdot\mathbf{n}^{T}-\mathbf{n}%
\cdot\mathbf{\pi}^{T}\right)  _{ij};
\end{align*}
in case b) we have
\begin{align*}
\mathcal{A}_{ij}  &  =0,\\
\mathcal{B}_{ij}  &  =\left(  I-\mathbf{n}\cdot\mathbf{n}^{T}\right)  _{ij};
\end{align*}
and, in case c), since the first term in (\ref{17p}) is antisymmetric in
$i\leftrightarrow j$ while the second term is not, we must require both terms
to vanish separately, yielding
\begin{align*}
\mathcal{C}_{ij}  &  =-\left(  \mathbf{\pi}\cdot\mathbf{n}^{T}-\mathbf{n}%
\cdot\mathbf{\pi}^{T}\right)  _{ij},\\
\mathcal{B}_{ij}  &  =\left(  I-\mathbf{n}\cdot\mathbf{n}^{T}\right)  _{ij}.
\end{align*}
It then follows from (\ref{16p}) that $\mathcal{A}_{ij}=2\left(
I-\mathbf{n}\cdot\mathbf{n}^{T}\right)  _{im}\left(  W^{-1}\right)
_{mj}\left(  \partial_{y}\right)  ^{-1}$, which is not acceptable because it
is not antisymmetric in $i\leftrightarrow j$. Therefore, we conclude that
there is no solution to the equations (\ref{16p}), (\ref{17p}) with $W=M^{-1}%
$. This implies that the AM boundaries are not compatible with the bulk $O(N)$
symmetry, which has been used to obtain the Poisson brackets (\ref{7}%
)-(\ref{9}) upon which the equations (\ref{16p}), (\ref{17p}) are based.
Therefore, we shall temporarily restrict ourselves to the cases a) and b).

By use of the equations (\ref{Jac:1})-(\ref{Jac:4}), we find that, for the
case a), i.e. AD boundaries, the following operators constitute a consistent
set of solution to (\ref{16})-(\ref{Jac:4}),%
\begin{align}
\mathcal{A}_{ij}  &  =0,\qquad\mathcal{B}_{ij}=-\left(  I-\mathbf{n}%
\cdot\mathbf{n}^{T}\right)  _{ij},\nonumber\\
\mathcal{C}_{ij}  &  =-\left(  \mathbf{\pi}\cdot\mathbf{n}^{T}-\mathbf{n}%
\cdot\mathbf{\pi}^{T}\right)  _{ij}. \label{sol:p0}%
\end{align}
For the case b), i.e. AN boundaries, the solution to (\ref{16})-(\ref{Jac:4})
is found to be%
\begin{align}
\mathcal{A}_{ij}  &  =0,\qquad\mathcal{B}_{ij}=\left(  I-\mathbf{n}%
\cdot\mathbf{n}^{T}\right)  _{ij},\nonumber\\
\mathcal{C}_{ij}  &  =\left(  \mathbf{\pi}\cdot\mathbf{n}^{T}-\mathbf{n}%
\cdot\mathbf{\pi}^{T}\right)  _{ij}. \label{sol:pN}%
\end{align}
Substituting the solutions (\ref{sol:p0}) and (\ref{sol:pN}) back into the
test Poisson brackets (\ref{11})-(\ref{13}), we get the following Poisson
brackets, which are consistent with AD and AN boundary conditions respectively
and satisfy all Jacobi identities simultaneously,%
\begin{align}
&  \,\{n_{i}(x),n_{j}(y)\}_{M}=0,\nonumber\\
&  \,\{n_{i}(x),\pi_{j}(y)\}_{M}\nonumber\\
&  \quad=(\delta_{ij}-n_{i}n_{j})\left[  \delta(x-y)-\delta(x+y)\right]
,\nonumber\\
&  \,\{\pi_{i}(x),\pi_{j}(y)\}_{M}\nonumber\\
&  \quad=(\pi_{i}n_{j}-n_{i}\pi_{j})\left[  \delta(x-y)-\delta(x+y)\right]  ,
\label{ADP}%
\end{align}%
\begin{align}
&  \,\{n_{i}(x),n_{j}(y)\}_{M}=0,\nonumber\\
&  \,\{n_{i}(x),\pi_{j}(y)\}_{M}\nonumber\\
&  \quad=(\delta_{ij}-n_{i}n_{j})\left[  \delta(x-y)+\delta(x+y)\right]
,\nonumber\\
&  \,\{\pi_{i}(x),\pi_{j}(y)\}_{M}\nonumber\\
&  \quad=(\pi_{i}n_{j}-n_{i}\pi_{j})\left[  \delta(x-y)+\delta(x+y)\right]  .
\label{ANP}%
\end{align}

The action (\ref{action}), together with the consistent Poisson brackets
(\ref{ADP}) (resp. (\ref{ANP})), form a complete Hamiltonian description for
classical $O(N)$ nonlinear sigma model in the presence of AD (resp. AN)
boundary conditions.

\subsection{The symmetry breaking boundary ND}

ND boundaries correspond to $1<p<N$ in (\ref{W}) and $\mathcal{W}$ diagonal
with $\mathcal{W}_{ii}\rightarrow\infty$ for all $i$. Since $O(N)$
transformations cannot transform Neumann boundary conditions into Dirichlet
ones, ND boundaries explicitly break the $O(N)$ symmetry into the subgroup
$O(p)\times O(N-p)$. Consequently, while considering the consistent
Hamiltonian description of the model in the presence of ND boundaries, we need
to modify not only the Poisson brackets at the boundary, but also in the bulk.
In fact, that the ND boundary conditions break not only the $O(N)$ symmetry at
the boundary but also in the bulk is an important conclusion of our study,
since it can be seen that the direct substitution of the $O(N)$ conditions
$\mathfrak{A}_{ij}=0,\mathfrak{B}_{ij}=\delta_{ij}-n_{i}n_{j}$ and
$\mathfrak{C}_{ij}=\pi_{i}n_{j}-\pi_{j}n_{i}$ together with the matrix $W$ in
(\ref{W})--with $\mathcal{W}$ diagonal and $\mathcal{W}_{ii}\rightarrow\infty$
for all $i$--into the equations (\ref{16}) and (\ref{17}) would lead to
contradictory results.

For convenience we divide the suffices $i,j$ etc of the fields into two
disjoint sets, labeled respectively by Latin and Greek letters. Latin indices
$a,b$ run from $1$ to $p$ and Greek indices $\alpha,\beta$ run from $p+1$ to
$N$. We also introduce the notations $\mathbf{n}^{(1)}=(n_{1},\cdots
,n_{p})^{T}$, $\mathbf{n}^{(2)}=(n_{p+1},\cdots,n_{N})^{T}$ and similarly
$\mathbf{\pi}^{(1)}=(\pi_{1},\cdots,\pi_{p})^{T}$, $\mathbf{\pi}^{(2)}%
=(\pi_{p+1},\cdots,\pi_{N})^{T}$. Then the $O(p)\times O(N-p)$ symmetric bulk
in the presence of ND boundaries can be described by the fields $\mathbf{n}%
^{(1)}$ and $\mathbf{n}^{(2)}$ obeying, respectively, $\mathbf{n}^{(1)T}%
\cdot\mathbf{n}^{(1)}=u,\mathbf{n}^{(2)T}\cdot\mathbf{n}^{(2)}=v$, where the
constants $u$ and $v$ satisfy $u+v=1$. The bulk Poisson brackets in this case
are characterized by (\ref{PBulk}) with the following functions
$\mathfrak{A,B}$ and $\mathfrak{C}$,%
\begin{align}
\mathfrak{A}_{ab}  &  =\mathfrak{A}_{a\beta}=\mathfrak{A}_{\alpha
b}=\mathfrak{A}_{\alpha\beta}=0,\nonumber\\
\mathfrak{B}_{ab}  &  =\delta_{ab}-n_{a}n_{b},\quad\mathfrak{B}_{a\beta
}=0,\nonumber\\
\mathfrak{B}_{\alpha b}  &  =0,\quad\mathfrak{B}_{\alpha\beta}=\delta
_{\alpha\beta}-n_{\alpha}n_{\beta},\nonumber\\
\mathfrak{C}_{ab}  &  =\pi_{a}n_{b}-\pi_{b}n_{a},\quad\mathfrak{C}_{a\beta
}=0,\nonumber\\
\mathfrak{C}_{\alpha b}  &  =0,\quad\mathfrak{C}_{\alpha\beta}=\pi_{\alpha
}n_{\beta}-\pi_{\beta}n_{\alpha}. \label{NDABC}%
\end{align}
Substituting (\ref{NDABC}) into (\ref{16}), (\ref{17}) and setting
$\mathcal{W}_{ij}=0$ for $i\neq j$ and $\mathcal{W}_{ii}\rightarrow\infty$ for
all $i$, we get, from (\ref{16})-(\ref{Jac:4}), the following consistent
solution,
\begin{align}
\mathcal{A}_{ab}  &  =0,\quad\mathcal{B}_{ab}=\delta_{ab}-n_{a}n_{b}%
,\nonumber\\
\mathcal{C}_{ab}  &  =\pi_{a}n_{b}-\pi_{b}n_{a},\nonumber\\
\mathcal{A}_{\alpha\beta}  &  =0,\quad\mathcal{B}_{\alpha\beta}=n_{\alpha
}n_{\beta}-\delta_{\alpha\beta},\nonumber\\
\mathcal{C}_{\alpha\beta}  &  =-\pi_{\alpha}n_{\beta}+\pi_{\beta}n_{\alpha
},\nonumber\\
\mathcal{A}_{a\beta}  &  =\mathcal{A}_{\alpha b}=\mathcal{B}_{a\beta
}=\mathcal{B}_{\alpha b}=\mathcal{C}_{a\beta}=\mathcal{C}_{\alpha b}=0.
\label{NDSol}%
\end{align}
The Poisson brackets (\ref{PBulk}) with $\mathfrak{A,B}$ and $\mathfrak{C}$
given in (\ref{NDABC}) and $\mathcal{A},\mathcal{B},\mathcal{C}$ in
(\ref{NDSol}) are nothing but the union of consistent Poisson brackets for an
$O(p)$ nonlinear sigma model with AN boundaries and those of an $O(N-p)$
nonlinear sigma model with AD boundaries, as they should be.

\subsection{The forbidden boundaries AM and MD}

That the AM boundaries are not compatible with the $O(N)$ symmetry in the bulk
has already been mentioned earlier in this section. This fact can also be seen
from another point of view. Following \cite{Moriconi3} and with a
straightforward generalization, we can see that the AM boundary conditions
(\ref{4pp}) with $W=M^{-1}$ can be realized on the lagrangian level by adding
to the bulk action (\ref{action}) with the boundary term
\begin{equation}
S_{b}=\left.  \int dtM_{ij}n_{i}\partial_{t}n_{j}\right\vert _{x=0}.
\label{Sb}%
\end{equation}
It can be easily seen that, under the global $O(N)$ transformation
$n_{i}\rightarrow O_{ij}n_{j}$, $M$ will transform as $M_{ij}\rightarrow
O_{ik}M_{kl}O_{lj}^{T}$. That $M$ does not commute with the generic element
$O$ of the group $O(N)$ is an explicit signature that the boundary term
(\ref{Sb}) is not invariant under $O(N)$. In fact, the maximal subgroup of
$O(N)$ which may leave the boundary term (\ref{Sb}) invariant is
$O(2)^{\otimes K}$, an abelian subgroup, in which case $M$ must be given in
the form of (\ref{M}). This explains our choice of $M$ in (\ref{M}).

Since the bulk $O(N)$ symmetry is broken by the AM boundary conditions into
$O(2)^{\otimes K}$, we may introduce the fields $\mathbf{n}^{(\ell)}%
=(n_{2\ell-1},n_{2\ell})^{T}$ and their conjugate momenta to describe the bulk
system as a union of $K$ $O(2)$ nonlinear sigma models, each obeys
$\mathbf{n}^{(\ell)T}\cdot\mathbf{n}^{(\ell)}=u_{\ell}$, with the constants
$u_{\ell}$ satisfying $\sum_{\ell=1}^{K}u_{\ell}=1$. Accordingly, the Poisson
brackets which are consistent in the bulk are just (\ref{PBulk}) with the
matrix functions $\mathfrak{A}$, $\mathfrak{B}$ and $\mathfrak{C}$ given,
respectively, by%
\begin{equation}
\mathfrak{A}=0,\quad\mathfrak{B=}%
{\textstyle\bigoplus\limits_{\ell=1}^{K}}
\mathfrak{B}^{(\ell)},\quad\mathfrak{C=}%
{\textstyle\bigoplus\limits_{\ell=1}^{K}}
\mathfrak{C}^{(\ell)}, \label{AMABC}%
\end{equation}
where $\mathfrak{B}^{(\ell)}$ and $\mathfrak{C}^{(\ell)}$ are all $2\times2$
matrices given as%
\begin{align}
\mathfrak{B}^{(\ell)}  &  =I_{2\times2}-\mathbf{n}^{(\ell)}\cdot
\mathbf{n}^{(\ell)T},\nonumber\\
\mathfrak{C}^{(\ell)}  &  =\mathbf{\pi}^{(\ell)}\cdot\mathbf{n}^{(\ell
)T}-\mathbf{n}^{(\ell)}\cdot\mathbf{\pi}^{(\ell)T}. \label{AMABCBL}%
\end{align}
Now substituting (\ref{AMABC}) and (\ref{AMABCBL}) into (\ref{16}) and
(\ref{17}), we get, at the $\ell$-th diagonal block, the following equations,
\begin{align}
&  \mathcal{A}_{im}W_{mj}^{(\ell)}\partial_{y}-\left(  I-\mathbf{n}^{(\ell
)}\cdot\mathbf{n}^{(\ell)T}+\mathcal{B}\right)  _{ij}=0,\label{AM:nosol1}\\
&  \left(  \mathbf{\pi}^{(\ell)}\cdot\mathbf{n}^{(\ell)T}-\mathbf{n}^{(\ell
)}\cdot\mathbf{\pi}^{(\ell)T}+\mathcal{C}\right)  _{ij}\nonumber\\
&  -W_{im}^{(\ell)}\left(  I-\mathbf{n}^{(\ell)}\cdot\mathbf{n}^{(\ell
)T}-\mathcal{B}\right)  _{mj}\partial_{y}=0, \label{AM:nosol}%
\end{align}
where $i,j=2\ell-1$ or $2\ell$, $W^{(\ell)}\ $is the $\ell$-th diagonal block
of $W$, which is given in (\ref{M}) through $W=M^{-1}$. It follows that there
is no solution to (\ref{AM:nosol1}) and (\ref{AM:nosol}), since the first term
in (\ref{AM:nosol1}) is diagonal, while the second term cannot be diagonal.
Similarly, the first term in (\ref{AM:nosol}) is anti-diagonal, but the second
term cannot be anti-diagonal.

Now we are forced to answer the following questions: What happens to the mixed
boundary conditions? Why couldn't we find any consistent Poisson brackets for
the $O(N)$ nonlinear sigma model in the presence of AM boundaries? Two
contradictory answers might be in order, which are 1) the AM boundaries are
completely incompatible with any orthogonal symmetry, i.e. even the $O(2)$'s
cannot survive after AM boundary conditions are applied; 2) the method we are
using to construct the consistent boundary Poisson brackets fails for the
mixed boundaries for $O(N)$ nonlinear sigma model. Our choice is the answer
1). To support our choice, we now consider the simplest case of $K=1$, i.e. a
single $O(2)$ nonlinear sigma model with mixed boundary conditions $\left(
\partial_{x}n_{i}+M_{ij}\partial_{t}n_{j}\right)  |_{x=0}=0,$ $M=g\left(
\begin{array}
[c]{cc}%
0 & -1\\
1 & 0
\end{array}
\right)  $. This is exactly the original boundary conditions studied in
\cite{Moriconi3,Moriconi2}. Expanding the above boundary conditions in
component form, we get%
\begin{align}
\left(  \partial_{x}n_{1}-g\partial_{t}n_{2}\right)  |_{x=0}  &
=0,\nonumber\\
\left(  \partial_{x}n_{2}+g\partial_{t}n_{1}\right)  |_{x=0}  &  =0.
\label{bound}%
\end{align}
On the other hand, from the $O(2)$ condition at the boundary, $\left(
n_{1}^{2}+n_{2}^{2}\right)  _{x=0}=1$, we can get%
\begin{align}
\left(  n_{1}\partial_{t}n_{1}+n_{2}\partial_{t}n_{2}\right)  |_{x=0}  &
=0,\label{symm:1}\\
\left(  n_{1}\partial_{x}n_{1}+n_{2}\partial_{x}n_{2}\right)  |_{x=0}  &  =0.
\label{symm:2}%
\end{align}
Substituting (\ref{bound}) into (\ref{symm:2}), it follows that%
\begin{equation}
\left(  n_{1}\partial_{t}n_{2}-n_{2}\partial_{t}n_{1}\right)  |_{x=0}=0.
\label{fin}%
\end{equation}
Combining (\ref{symm:1}) and (\ref{fin}) with the $O(2)$ condition $\left(
n_{1}^{2}+n_{2}^{2}\right)  _{x=0}=1$, we get both $\partial_{x}n_{i}%
|_{x=0}=0$ and $\partial_{t}n_{i}|_{x=0}=0$. In other words, if the mixed
boundaries are applied, the fields $n_{i}$ will obey both Neumann and
Dirichlet boundary conditions simultaneously. This is certainly impossible, so
we end up with the surprising conclusion that the mixed boundaries are
actually not allowed in $O(N)$ nonlinear sigma model, not to say their
integrability. This conclusion removes the AM as well as MD boundary
conditions from the allowed list of integrable boundaries.

\section{Discussions}

Using the method proposed in \cite{Zhao} and developed in \cite{He} and
\cite{Zhao2}, we analyzed the problem of consistent Poisson brackets for
classical $O(N)$ nonlinear sigma model in the presence of five different sets
of boundary conditions, i.e. the AD, AN, ND, AM and MD boundaries. Only in the
presence of AD, AN and ND boundaries we have found consistent Poisson
brackets, while for AM and MD boundaries, no consistent Poisson brackets can
be found, showing that the mixed boundary conditions are completely
incompatible with any orthogonal symmetry.

Through the analysis of ND boundaries, we find that the idea underlying our
method needs a significant modification. The original statement that in the
presence of boundary constraints the Poisson brackets need to be modified only
at the boundary is only valid if the boundary conditions preserve all the bulk
symmetries. On the other hand, if the boundary conditions are symmetry
breaking, they will also affect the bulk part of the Poisson brackets, so that
the final consistent Poisson brackets have the same symmetry in the bulk and
at the boundary.

The result of this Letter not only widens the scope of applicability of the
method of \cite{Zhao}, but also has important applications in the study of
$O(N)$ nonlinear sigma model itself. A straightforward application might be in
the study of current algebra in the presence of boundary conditions, which is
an important ingredient in the classical integrable structure of the model.
For instance, the Poisson algebra calculations made in \cite{Corrigan3} should
be re-examined using our result (\ref{ANP}), because the bulk Poisson brackets
(\ref{7})-(\ref{9}) are no longer consistent in the presence of Neumann
boundaries as used in \cite{Corrigan3}.

\section*{Acknowledgement}

This work is supported in part by the National Natural Science Foundation of China.

\bibliographystyle{utcaps}
\bibliography{Zhao-He}

\end{document}